# COVID-19 Pandemic Severity, Lockdown Regimes, and People's Mobility: Early Evidence from 88 Countries

**Md. Mokhlesur Rahman [1,2], Jean-Claude Thill [3,\*] and Kamal Chandra Paul [4]**

[1] Department of Urban and Regional Planning, Khulna University of Engineering & Technology, Khulna-9203, Bangladesh
[2] INES Program, The William States Lee College of Engineering, The University of North Carolina at Charlotte, 9201 University City Blvd, Charlotte, NC 28223, USA; mrahma12@uncc.edu
[3] Department of Geography and Earth Sciences and School of Data Science, The University of North Carolina at Charlotte, 9201 University City Blvd, Charlotte, NC 28223, USA
[4] Department of Electrical and Computer Engineering, The William States Lee College of Engineering, The University of North Carolina at Charlotte, 9201 University City Blvd, Charlotte, NC 28223, USA; kpaul9@uncc.edu
\* Correspondence: jfthill@uncc.edu; Tel.: +1-704-687-5973

**Abstract:** This study empirically investigates the complex interplay between the severity of the coronavirus pandemic, mobility changes in retail and recreation, transit stations, workplaces, and residential areas, and lockdown measures in 88 countries around the world during the early phase of the pandemic. To conduct the study, data on mobility patterns, socioeconomic and demographic characteristics of people, lockdown measures, and coronavirus pandemic were collected from multiple sources (e.g., Google, UNDP, UN, BBC, Oxford University, Worldometer). A Structural Equation Modeling (SEM) framework is used to investigate the direct and indirect effects of independent variables on dependent variables considering the intervening effects of mediators. Results show that lockdown measures have significant effects to encourage people to maintain social distancing so as to reduce the risk of infection. However, pandemic severity and socioeconomic and institutional factors have limited effects to sustain social distancing practice. The results also explain that socioeconomic and institutional factors of urbanity and modernity have significant effects on pandemic severity. Countries with a higher number of elderly people, employment in the service sector, and higher globalization trend are the worst victims of the coronavirus pandemic (e.g., USA, UK, Italy, and Spain). Social distancing measures are reasonably effective at tempering the severity of the pandemic.

## 1. Introduction

The novel coronavirus, also known as Coronavirus Disease 2019 (COVID-19), first emerged in Wuhan (P.R. China) in late fall 2019 and has now spread to 213 countries around the globe [1]. The World Health Organization (WHO) declared COVID-19 a pandemic on 11 March 2020, considering its outbreak in many countries [2]. As of now, more than 44 million people have been infected by this highly infectious disease and over 1.1 million people have died [1,3]. The current fatality rate among closed cases is about 10%, which speaks volume about the sheer severity of the pandemic. The increasing number of coronavirus cases and deaths poses challenges to the healthcare system, economic development, supply chain, education, and travel pattern of the people [4]. Figure 1 represents the increasing number of coronavirus infection cases and deaths (in million) in the world from 31 December 2019 to 29 October 2020. To control the spread of COVID-19, governments have implemented travel bans through national lockdown, stay-at-home order, restriction on mass gathering and non-essential travel, which further affected people's mobility and social distancing practices.

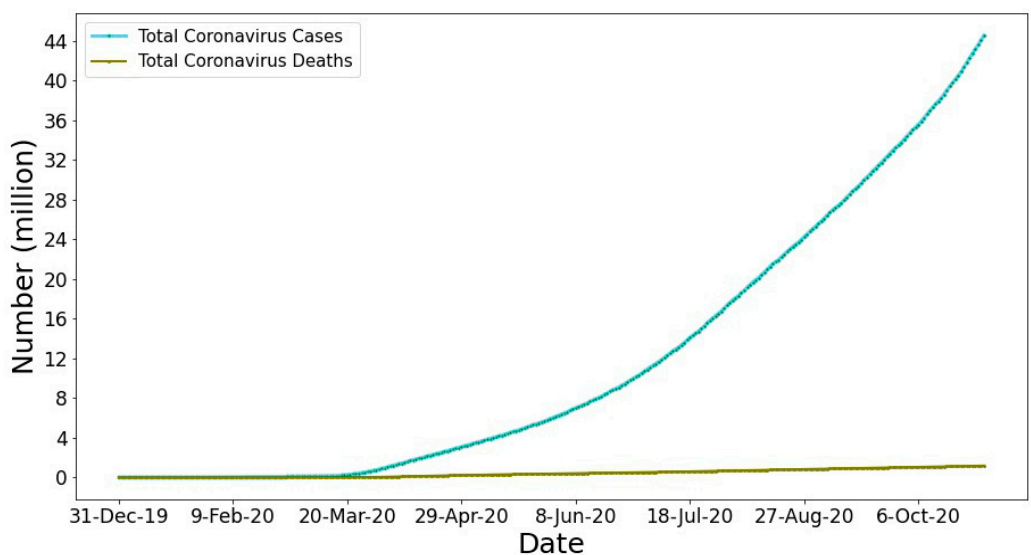

**Figure 1.** Total number of coronavirus infection cases and deaths in the world [5].

This study mainly aims at unraveling the complex relationships between the incidence of the pandemic, lockdown measures on populations and their social distancing and mobility behaviors in the early stage of the pandemic. The following three research questions are form the core of the understanding these interrelated relationships:

(1) What are the impacts of different lockdown measures on reducing people's mobility patterns and the severity of the pandemic?
(2) What are the effects of socioeconomic and institutional arrangements and dispositions on population mobility and on the pandemic severity?
(3) What are the consequences of pandemic severity on the condition of social distancing of the people?

The impacts of COVID-19 on public health have been discussed in many previous papers [6–9]. This disease is imposing tremendous pressure on the health care system [8]. Besides, COVID-19 is affecting the mental health of people in the form of mass fear, panic, and uncertainties [10–12]. Because of the escalation of the pandemic, there has been a huge increase in the personal stockpiling of necessary goods (e.g., food staples, toilet paper, cleaning supply) which is unsettling the balance in the demand and supply of consumer goods [6].

Many researchers have investigated the impacts of COVID-19 on the global economy [4,6,9,13,14]. Globally, stock markets collapsed by 50%. As COVID-19 threw millions out of work in the US, it caused an unemployment rate soaring to 14.7% in April 2020, which is the highest rate since the Great Depression [15,16]. The US Congress passed a $2 trillion coronavirus aid package to help businesses and workers. Global annual GDP is expected to contract by 3–4%. With the COVID-19 outbreak, a massive freeze in the industrial and logistical infrastructure caused a devastation throughout the global economy. Many investors moved towards safer investments because of the fear of a worldwide recession [9]. Meanwhile, the global supply chain has been deeply disrupted. About 940 of the Fortune 1000 companies have reported a supply chain disruption due to COVID-19 [17]. A simulation study observed that changes in opening and closing time of the facilities due to the coronavirus pandemic are affecting supply chain performance [18]. However, considering the sharp

economic downturn, people are also very much concerned about reopening the economy. A recent study using Twitter data indicated that Americans are more supportive than fearful regarding reopening the economy [11]. Thus, adequate protective measures need to be adopted to safeguard people from COVID-19, even if the authorities forge ahead with a normal reopening of the economy.

Meanwhile, the travel industry is now facing an unprecedented reduction of flights, both internationally and domestically [14] after years of unbridled growth. As a precautionary measure in the face of the outbreak, human mobility has been curtailed across the board, entailing reductions in long-distance travel as well as in household trips for daily activities. This is an indirect consequence of the pandemic, which the world previously experienced during the Severe Acute Respiratory Syndrome (SARS) and the Middle East Respiratory Syndrome (MERS) outbreaks of 2002–2003 and 2011–2012, respectively. The virus has spread fast because of the transmission from infected regions to uninfected regions through the movement of people [7]. The analysis of mobility-based data suggested that a simultaneous restriction on travel across different regions and migration control is an effective way to control the spread of the virus [19–22]. Additionally, constrained human mobility by enacting lockdown or shelter-in-place orders can control community transmission of the virus.

The outflow of population from the infected regions poses a major threat to the destination regions. Mass transport (e.g., buses, trains) plays a very important role in the importation of COVID-19. A positive correlation of case importation has been found with the frequency of flights, buses, and trains from infected cities [20]. Thus, travel from the infected cities and regions can reduce the rapid transmission of the COVID-19. Similarly, different non-pharmaceutical interventions (NPIs) (e.g., travel ban, school, and public transport closure, restriction on public gathering, stay-at-home order) imposed by governments can mitigate community transmission of the COVID-19 in the affected regions, which dramatically curtails the mobility of people [6,13,21,23–33].

Apart from essential trips, non-essential businesses, amusement parks, cinemas, sports venues, public events, and exhibitions are curtailed. Nowadays, people are adjusting their travel decisions voluntarily to avoid coronavirus infection. Moreover, people are canceling and postponing their trips because of perceived danger and negative impacts on the health of family members and relatives [34]. A recent study using GPS location-based data observed that a change in infection rate of 0.003% is accompanied by mobility reduction in the order of 2.31% at the county level in the US [35]. On the other hand, the stay-at-home order reduces mobility by 7.87%. Thus, lockdown measures are very effective means of social distancing and ultimately alleviating pandemic severity. This study also observed higher mobility reduction in the counties with a higher number of elderly people, lower share of republican supporters in the 2016 presidential election, and higher population density.

Travel bans and restrictions provide some reprieve that is very helpful to reinforce and establish necessary measures in controlling the spread of the epidemic [33]. Researchers estimated that travel reduction from 28 January to 07 February 2020 prevented 70.4% coronavirus infections in China [26]. Using the count data model they observed that travel restriction resulted in the delay of a major epidemic by two days in Japan, and the probability of a major epidemic reduced by 7 to 20%. Researchers in [36] developed an interactive web map to show the spatial variation of mobility during the COVID-19 pandemic. Analyzing county-level mobility data released by SafeGraph, this study found that mobility decreased considerably by March 31, 2020 in the US, when most states ordered lockdown and imposed stay-at-home orders. Using the susceptible-exposed-infectious-recovered (SEIR) model, studies in Taiwan [27] and Europe [31] showed that reduction of intercity and air travel, respectively, can effectively reduce the coronavirus pandemic. However, using the same methodology, another study commented that travel restriction may be an effective measure for a short term case, yet it is ineffective to eradicate the disease as it is impossible to remove the risk of seeding the virus to other areas [25].

National and international travel restrictions may only modestly delay the spread of the virus unless there is a certain level of control in community transmission (i.e., inability to identify the sources of infections). Using a global metapopulation disease transmission model, researchers observed that even with 90% travel restrictions to and from China, only a mild reduction in coronavirus pandemic could be envisioned until community transmission is reduced by 50% at least [28]. Thus, appropriate NPIs to reduce community transmission are necessary to weaken the pandemic. Similarly, pharmaceutical interventions (PIs) are mandatory to provide proper medication to infected people and improve health conditions. Thus, a coordinated effort comprising NPIs and PIs is necessary to mitigate the impacts of COVID-19 [37].

Reduction in community transmission is seen as an effective measure to control coronavirus severity. On the other hand, lockdown regimes such as local travel ban, stay-at-home order, restrictions on public gatherings, and school closures, essentially reduce community transmission of the COVID-19 by reducing the mobility of the people. Because there is no theoretical basis to hold the view these are simple dependencies, this study assesses how lockdown measures on populations, their social distancing and mobility behaviors, and the severity of the COVID-19 pandemic triangulate to portray the public health state of a country. Also, we study how the socioeconomic and institutional contexts of a country condition the specific modalities of these relationships. The analysis is conducted within the framework of a Structural Equation Model (SEM).

Based on the literature review, a conceptual framework has been developed (Figure 2). The conceptual framework posits that socioeconomic and institutional contexts have a significant role in pandemic severity, social distancing, and in the enactment of lockdown measures. Different lockdown measures implemented in affected countries influence pandemic severity and social distancing (i.e., mobility). Moreover, lockdown measures indirectly influence pandemic severity by changing people's mobility. Social distancing has a direct effect on pandemic severity. A high level of social distancing (i.e., reduction of mobility) is considered an effective measure to reduce infectious diseases. However, pandemic severity also has a direct effect on how people effectively practice social distancing, which implies that self-motivated people reduce their mobility when the severity of the pandemic is higher.

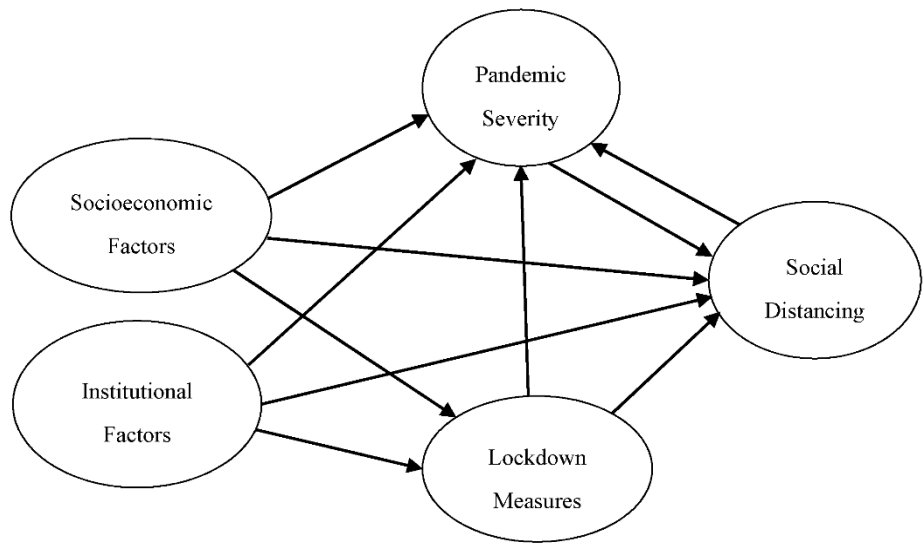

**Figure 2.** Conceptualization of the study.

This study empirically explores the relationships between lockdown measures, mobility patterns, pandemic severity, and socioeconomic and institutional factors of 88 countries in the world using SEM. Collecting data from multiple sources, this study finds that lockdown measures have significant influence on reducing mobility and thus control the severity of the pandemic. Social distancing also has direct impact on reducing pandemic severity, although the effect is rather marginal. The socioeconomic and institutional factors of a country significantly affect pandemic severity. However, pandemic severity, socioeconomic and institutional factors have no significant impacts on social distancing.

## 2. Materials and Methods

*2.1. Data*

To test and validate the conceptual model in Figure 2, data were collected from multiple sources (Table 1). The data sources include Google, the United Nations (UN), United Nations Development Program (UNDP), Worldometer, Oxford University, Hofstede, The Fraser Institute, KOF Swiss Economic Institute, and BBC.

**Table 1.** Description of the variables and data sources.

| Variable | Description | Source |
| --- | --- | --- |
| RR | Percentage change of mobility in retail and recreation trips | [38] |
| TS | Percentage change of mobility in transit stations trips | [38] |
| WP | Percentage change of mobility in workplaces trips | [38] |
| RD | Percentage change of mobility in residential trips | [38] |
| l_case | Total coronavirus infection cases per 1 million population | [1] |
| l_death | Total coronavirus deaths per 1 million population | [1] |
| NL | National lockdown | [39] |
| WPC | Workplace closing | [40] |
| SH | Stay-at-home order | [40] |
| SI | Stringency index [1] | [40] |
| FS | Percentage of female smokers | [40] |
| AGE65 | Percentage of the population age 65 and older | [40] |
| MA | Median age | [41] |
| EI | Average of years of schooling vs. expected years of schooling | [41] |
| AE | Percentage of the population employed in agriculture | [42] |
| SE | Percentage of the population employed in services | [42] |
| HE | Percentage of health expenditure to total GDP | [42] |
| IDV | Individualism versus Collectivism emphasis [2] | [43] |
| KOFGI | KOF Globalization Index [3] | [44] |

[1] A composite index considering all implemented lockdown measures. The score of this index ranges from 0 to 100. A high score indicates the strictest measures and low score indicates loose measure. [2] This indicator measures the degree of interdependence among the members of a society. The score ranges from 0 to 100. A low score indicates collective culture and higher interdependence among the members and conversely a high score indicates Individualist culture and a low level of interdependence. [3] A composite index that indicates openness to trade and capital flows considering economic, social and political aspects. The score of the index ranges from 0 to 100. A high score denotes a highly globalized country and a low score indicates poorly globalized country.

Google prepared a COVID-19 Community Mobility Report to help policymakers and public health professionals to understand changes in mobility in responses to lockdown measures (e.g., travel ban, work-from-home, shelter-in-place, restriction on public gathering) [38,45]. This report shows how visits and length of stay at different places, such as retail and recreation (e.g., restaurant, café, shopping center, theme park), workplaces (i.e., place of work), transit stations (e.g., subway stations, seaport, taxi stand, rest area), residential areas (i.e., place of residence), parks (e.g., public park, national forest), grocery stores and pharmacies (e.g., supermarket, convenience store, drug store) changed as of April 17 compared to a baseline value, with a potential to reduce the impact of COVID-19 pandemic. The baseline value is the median value of the corresponding week during the 5-week period from 3 January to 6 February 2020. The data were collected from the Google account holders who have turned on their travel location history. This study uses mobility changes in retail and recreation, workplaces, transit stations, and residential areas for 88 countries (Figure 3). Due to the ambiguity of the nature of visits and trips to grocery stores and pharmacies and the inconsistent definition of parks across countries (i.e., only include public parks), mobility changes in these two points of interest (POIs) were excluded from the study.

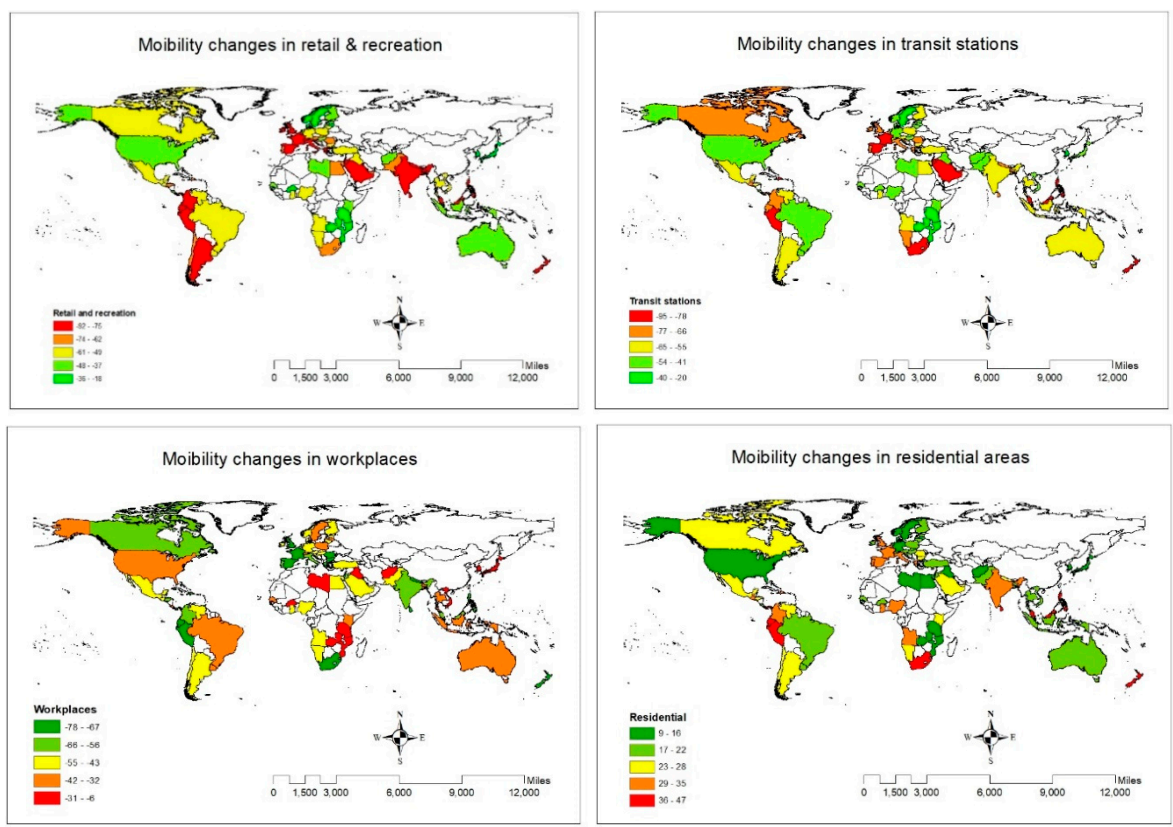

**Figure 3.** Mobility changes in POIs.

The total number of coronavirus infection cases and death cases as of April 17 were collected from Worldometer [1]. They collect data from thousands of sources around the world, analyze and validate them in real-time, and provide COVID-19 live statistics. To flatten the curve of COVID-19, governments issued different lockdown measures for part or whole country to restrict all non-essential movements. Data related to lockdown measures were collected from Dunford, Dale [39] and Oxford [40]. This study also collected socioeconomic (e.g., age, education, employment sector) and institutional context (e.g., individualism versus collectivism, globalization index) data to investigate their impacts on coronavirus infection cases and deaths, lockdown measures, and travel patterns (Table 1). After collecting data for 88 countries, they were integrated to build a complete dataset and conduct this study. Table 1 lists the variables that were included in the final model. A complete list of variables that were tested in the SEM framework to achieve the final model is provided in Appendix A.

Descriptive statistics of 19 different social distancing measures, lockdown variables, coronavirus infection cases and deaths, socioeconomic, and institutional context variables of all 88 countries are reported in Table 2. They are included in the statistical model as dependent variables, independent variables, mediators, and control variables.

**Table 2.** Descriptive statistics of the variables (N = 88).

| Variable | Unit | Min | Max | Mean | SD |
|---|---|---|---|---|---|
| RR | % | −92 | −18 | −59.41 | 18.20 |
| TS | % | −95 | −20 | −60.91 | 15.06 |
| WP | % | −78 | −6 | −48.41 | 16.86 |
| RD | % | 7 | 47 | 24.19 | 8.78 |
| l_case | Cases/1M pop | 0.69 | 8.64 | 4.98 | 2.10 |
| l_death | Deaths/1M pop | 0.69 | 6.11 | 2.25 | 1.52 |
| NL | Dummy (1, 0) | 0 | 1 | 0.59 | 0.49 |
| WPC | Dummy (1, 0) | 0 | 1 | 0.83 | 0.38 |
| SH | Dummy (1, 0) | 0 | 1 | 0.67 | 0.47 |
| SI | Index | 38.22 | 100 | 82.07 | 13.73 |
| FS | % | 0.2 | 35.3 | 13.02 | 10.05 |
| AGE65 | % | 1.14 | 27.05 | 11.09 | 6.88 |
| MA | Year | 16.7 | 48.4 | 33.68 | 8.95 |
| EI | Index | 0.3 | 1 | 0.72 | 0.16 |
| AE | % | 0.1 | 73.2 | 16.72 | 18.51 |
| SE | % | 21.1 | 87.6 | 61.79 | 15.83 |
| HE | % | 2.4 | 17.1 | 6.95 | 2.74 |
| IDV | Index | 6 | 91 | 40.02 | 22.95 |
| KOFGI | Index | 38.2 | 91.3 | 71.82 | 12.96 |

*2.2. Statistical Model*

SEM is used to investigate the causal relationships between socioeconomic and institutional factors, lockdown variables, coronavirus infection and death rates, and social distancing measures. This multivariate statistical technique is a common method to investigate complex relationships between dependent variables, independent variables, mediators, and latent dimensions. Many researchers have used SEM to investigate the factors that affect travel behaviors (e.g., mode choice, trip purpose, travel distance), for instance [46–49]. SEM consists of regression analysis, factor analysis, and path analysis to explore interrelationships between variables. It is a confirmatory technique where an analyst tests a model to check consistency between the existing theories and the nature of constructs.

Based on Exploratory Factor Analysis (EFA) and extant theories, latent dimensions are created to reduce dimensions and easily understand the data and represent underlying concepts. The following four latent dimensions are created based on the observed data:

(1) Social distancing measures: TS, RR, WP, and RD
(2) Pandemic severity: l_case and l_death
(3) Lockdown measures: NL, WPC, SH, and SI
(4) Socioeconomics and institutional factors: MA, AGE65, KOFGI, AE, SE, HE, FS, EI, and IDV

Moreover, a path diagram is constructed to graphically represent interdependencies of the independent variables, mediators, and dependent variables in the model specification. Finally, a set of fit indices (e.g., Chi-square, CFI, TLI, RMSEA, SRMR) are estimated to establish goodness-of-fit of the model.

**3. Results**

*3.1. Calibrated Model*

The model is calibrated using the SEM Builder within STATA 15 [50]. The maximum likelihood estimation method is used to calculate the coefficients. The overall structure of the model with direct standardized coefficients is depicted in Figure 4. The final structure of the model includes interactions between dependent and independent variables through mediators.

The fit of the calibrated model is evaluated based on several goodness-of-fit statistics (Table 3). The Chi-square statistic of the estimated model is 261.331. A lower value of the Chi-square indicates a better-fit model. Other fit statistics confirm that the estimated model is satisfactory. Thus, by all accounts, the goodness-of-fit of the estimated SEM is within the acceptable range and is quite satisfactory, which validates the use of this model [46,47].

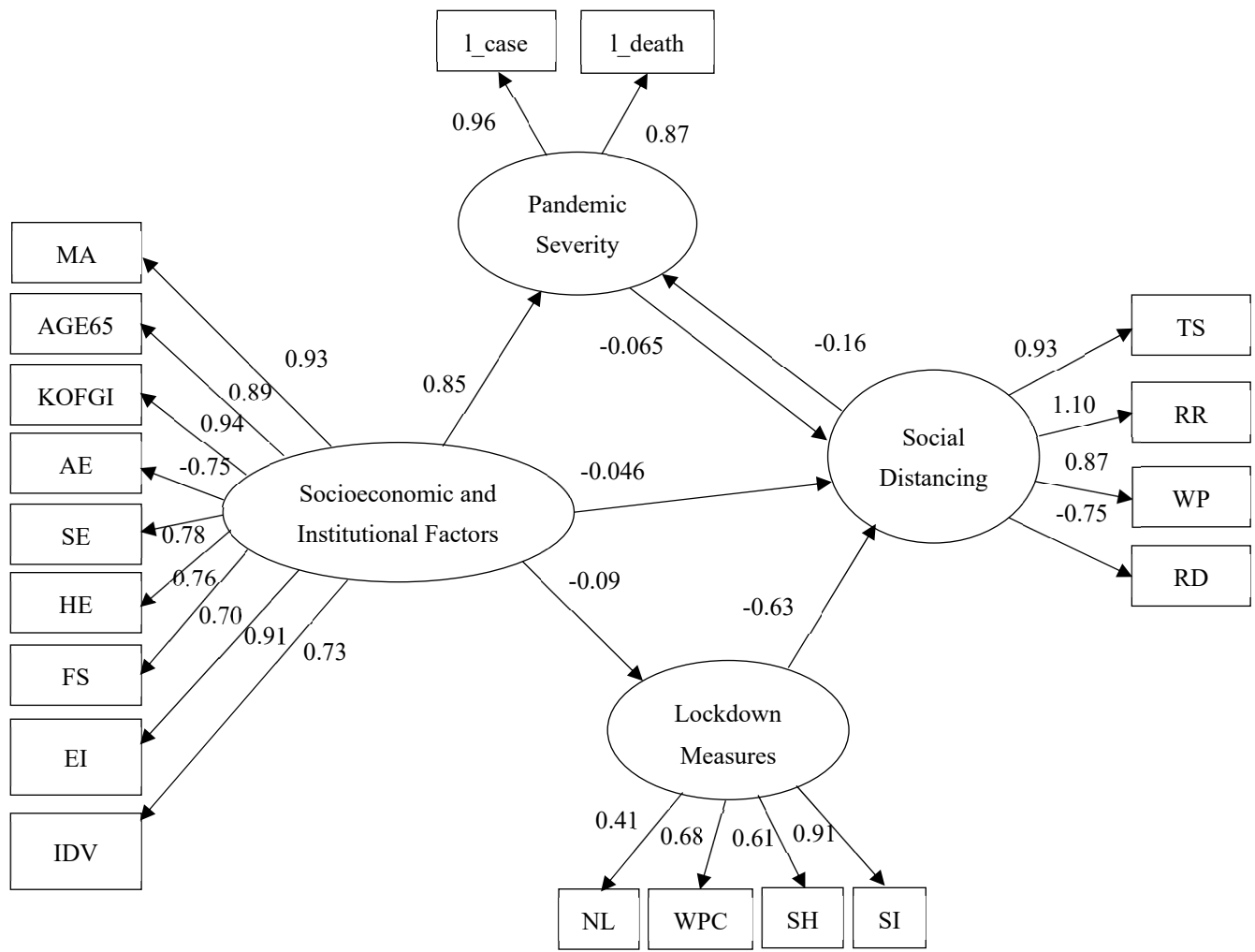

**Figure 4.** The calibrated model with direct standardized effects.

**Table 3.** Goodness-of-fit Statistics.

| Fit Statistic | Value |
| --- | --- |
| Chi-square | 261.331 |
| Chisq/df | 2.026 |
| RMSEA (Root mean squared error of approximation) | 0.108 |
| CFI (Comparative fit index) | 0.920 |
| TLI (Tucker-Lewis index) | 0.894 |
| SRMR (Standardized root mean squared residual) | 0.099 |

*3.2. Standardized Direct Effects*

Table 4 reports on the standardized coefficients by pair of variables in the model, including the direction of the modeled effect. These coefficients indicate the direct impacts of the socioeconomic and institutional factors, on the dependent variables of lockdown measures, pandemic severity and social distancing measures, and the direct interactions between and among dependent, independent, and latent variables. However, this table does not represent any indirect effects of independent variables through mediators. Table 4 also reports the standard error, z-value, and probability level (P-value) of the estimates. Most of the coefficients are statistically significant at the 0.001 level. However, a few coefficients with a P-value greater than 0.001 are retained in the model to preserve the overall representation of the relationships.

Table 4. Estimated standardized direct effects (N = 88).

| Variables | Std. Coef. | Std. Err. | z | P > z |
|---|---|---|---|---|
| Social Distancing <----------- Pandemic severity | −0.065 | 0.220 | −0.300 | 0.767 |
| Social Distancing <----------- Lockdown measures | −0.626 | 0.093 | −6.760 | 0.000 |
| Social Distancing <----------- Socioeconomic & institutional | −0.046 | 0.204 | −0.230 | 0.821 |
| Pandemic severity <---------- Social distancing | −0.160 | 0.086 | −1.860 | 0.063 |
| Pandemic severity <--------- Socioeconomic & institutional | 0.847 | 0.037 | 22.680 | 0.000 |
| Lockdown measures <------ Socioeconomic & institutional | −0.090 | 0.115 | −0.790 | 0.432 |
| MA <----------- Socioeconomic & institutional | 0.925 | 0.020 | 46.660 | 0.000 |
| IDV <----------- Socioeconomic & institutional | 0.734 | 0.050 | 14.720 | 0.000 |
| HE <----------- Socioeconomic & institutional | 0.762 | 0.053 | 14.400 | 0.000 |
| FS <----------- Socioeconomic & institutional | 0.701 | 0.054 | 12.910 | 0.000 |
| EI <----------- Socioeconomic & institutional | 0.911 | 0.021 | 43.400 | 0.000 |
| SE <----------- Socioeconomic & institutional | 0.777 | 0.046 | 16.920 | 0.000 |
| AE <----------- Socioeconomic & institutional | −0.746 | 0.048 | −15.400 | 0.000 |
| AGE65 <----------- Socioeconomic & institutional | 0.887 | 0.023 | 38.130 | 0.000 |
| KOFGI <----------- Socioeconomic & institutional | 0.936 | 0.017 | 54.230 | 0.000 |
| RR <----------- Social distancing | 1.133 | 0.109 | 10.380 | 0.000 |
| TS <----------- Social distancing | 0.933 | 0.038 | 24.760 | 0.000 |
| WP <----------- Social distancing | 0.868 | 0.046 | 18.800 | 0.000 |
| RD <----------- Social distancing | −0.751 | 0.054 | −13.980 | 0.000 |
| l_case <----------- Pandemic severity | 0.963 | 0.021 | 45.320 | 0.000 |
| l_death <----------- Pandemic severity | 0.872 | 0.031 | 28.310 | 0.000 |
| NL <----------- Lockdown measures | 0.413 | 0.099 | 4.180 | 0.000 |
| WPC <----------- Lockdown measures | 0.681 | 0.066 | 10.390 | 0.000 |
| SH <----------- Lockdown measures | 0.614 | 0.076 | 8.120 | 0.000 |
| SI <----------- Lockdown measures | 0.911 | 0.047 | 19.390 | 0.000 |

Four latent dimensions are created to understand social distancing, pandemic severity, lockdown measures, and socioeconomic and institutional characteristics. Now we discuss the model results by observing the relationships between latent dimensions and observed independent variables:

*Social Distancing*: This latent dimension is created from four observed variables: TS, RR, WP, and RD. It is the only dependent latent factor that represents the level of mobility changes of the people at transit stations, retail and recreation facilities, workplaces, and residences. Social distancing is positively associated with changes in the use of transit stations (0.933), retail and recreation facilities (1.133), and workplaces (0.868). In contrast, social distancing is negatively associated with residences (−0.751). Moreover, social distancing is negatively associated with pandemic severity (−0.160). All other things being held equal, a one-unit change in social distancing reduces pandemic severity by 0.16 units by reducing people's mobility, and consequently the risk of exposure to other individuals infected by the COVID-19 virus. Thus, increasing social distancing reduces the severity of the coronavirus pandemic (i.e., number of infection cases, and deaths). However, the relationship is marginally significant at a P-value of 0.063.

*Pandemic Severity*: This endogenous latent dimension is measured by two observed variables: l_case and l_death. Pandemic severity is positively associated with both of the observed variables (l_case: 0.963 and l_death: 0.876). In contrast, pandemic severity is negatively associated with social distancing, which implies that increasing severity of the pandemic reduces mobility in transit stations, retail and recreation, and workplaces and increases the time spent at one's residence. However, the association is not statistically significant (*P*-value: 0.767).

*Lockdown measures*: This endogenous latent factor is estimated by using four observed variables: NL, WPC, SH, and SI. The Lockdown latent dimension is positively associated with all of the measures (NL: 0.413, WPC: 0.681, SH: 0.614, and SI: 0.911) taken by governments to bring about social distancing and control the pandemic. Furthermore, lockdown measures are negatively associated with social distancing (−0.626). Thus, adopting strict lockdown measures (e.g., restriction on public gathering, workplace closing, and stay-at-home order) significantly reduces mobility at transit stations, retail and recreation facilities, and workplaces and increases time spent near one's home, all of which entailing people to stay home and avoid unnecessary travel.

*Socioeconomic and institutional factors*: This is the only exogenous latent dimension in the model. It comprises nine observed variables: MA, AGE65, KOFGI, AE, SE, HE, FS, EI, and IDV. Socioeconomics and institutional factors are positively associated with median age (0.925), elderly people (0.887), level of globalization (0.936), employment in the service sector (0.777), expenditure on health (0.762), female smokers (0.701), level of education (0.911), and the degree of interdependence in the society (0.734). Conversely, it is negatively associated with employment in the agricultural sector (−0.746). This latent dimension can therefore be interpreted as an indicator of urbanity and modernity. Moreover, socioeconomic and institutional factors are positively associated with pandemic severity (0.847) and negatively associated with lockdown measures (−0.090) and social distancing (−0.046). Thus, one unit change in socioeconomic and institutional factors leads to an increase in pandemic severity by 0.847 unit through increases in the number of elderly people, level of globalization, employment in the service sector, and reduction in employment in the agricultural sector. In contrast, one unit change in socioeconomic and institutional factors lead to a decrease in lockdown measures and in social distancing by 0.090 units and 0.046 units, respectively, by encouraging people to be more considerate of their impact on the rest of society. However, the impacts of socioeconomic and institutional factors on lockdown measures and social distancing are very minor and statistically non-significant at P-value 0.05.

### 3.3. Estimated Standardized Total Effects

It is important to analyze the total effect of latent factors on social distancing, pandemic severity, and lockdown measures considering their indirect effects, which remain unrevealed in the path diagram (Figure 4). Table 5 details the standardized total effects of latent factors on each of the observed variables of social distancing, pandemic severity, and lockdown regime.

Table 5. Total effects on social distancing and pandemic severity.

| Latent Factor | Social Distancing | | | | Pandemic Severity | | Lockdown Measures | | | |
|---|---|---|---|---|---|---|---|---|---|---|
| | TS | RR | WP | RD | l_case | l_death | NL | WPC | SH | SI |
| Pandemic severity | −0.061 | −0.075 | −0.057 | 0.049 | - | - | - | - | - | - |
| Lockdown measures | −0.591 | −0.717 | −0.550 | 0.475 | 0.098 | 0.088 | - | - | - | - |
| Socioeconomic and institutional factors | −0.042 | −0.051 | −0.039 | 0.034 | 0.823 | 0.746 | −0.037 | −0.061 | −0.055 | −0.082 |
| Social distancing | - | - | - | - | −0.156 | −0.141 | - | - | - | - |

Taking into account both direct and indirect effects, the analysis reveals that pandemic severity, lockdown measures, and socioeconomic and institutional factors reduce mobility at transit stations, retail and recreation centers, and workplaces and increase residential mobility. However, lockdown measures have much stronger and significant effects on all four social distancing aspects than pandemic severity and socioeconomic and institutional factors. In addition, the SEM analysis shows that lockdown and socioeconomic and institutional factors magnify pandemic severity while social distancing reduces pandemic severity. However, the impacts of socioeconomic and institutional factors are higher and statistically significant than lockdown measures and social distancing. Thus, lockdown measures are important to persuade people to stay home and maintain social isolation, and socioeconomic and institutional variables of urbanity and modernity substantially increase the severity of coronavirus pandemic. The table also indicates that only socioeconomic and institutional factors have direct impacts on the lockdown regime. However, the impacts are very insignificant.

Considering the complex relationships on hand, SEM extracts direct and indirect effects of variables and latent dimensions on social distancing, pandemic severity, and lockdown regime (Table 6). Direct and indirect impacts allow us to comprehend the core causes of social distancing and pandemic severity in different countries. Observing the direct and indirect effects, we understand that the direct effects of different latent factors on social distancing and pandemic severity is higher and significant compared to indirect effects. In some cases, indirect effects are statistically insignificant and often trivially small (Table 6). Thus, overall, the direct effects of latent factors represent the total effects without any mitigating or amplifying indirect effects. Results articulated in Table 6 illustrates that lockdown measures directly reduce people's mobility while socioeconomic and institutional factors increase the severity of the pandemic to a greater extent. Socioeconomic and institutional factors only have direct effects on lockdown, without any indirect influence on it.

**Table 6.** Direct, indirect, and total effects on social distancing, pandemic severity, and lockdown regime.

| Latent Factor | Social Distancing | | | Pandemic Severity | | | Lockdown | | |
|---|---|---|---|---|---|---|---|---|---|
| | Direct | Indirect | Total | Direct | Indirect | Total | Direct | Indirect | Total |
| Pandemic severity | −0.065 | −0.001 | −0.066 | - | 0.011 | 0.011 | - | - | - |
| Lockdown measures | −0.626 | −0.007 | −0.633 | - | −0.101 | −0.101 | - | - | - |
| Socioeconomic and institutional factors | −0.046 | 0.001 | −0.045 | 0.847 | 0.007 | 0.855 | −0.090 | - | −0.090 |
| Social distancing | - | 0.011 | 0.011 | −0.160 | −0.002 | −0.162 | - | - | - |

## 4. Discussion and Conclusions

COVID-19 has become a piercing issue and its numerous negative impacts on public health, economy, lifestyle, and wellbeing of populations are striking policymakers to come up with some solutions. To this end, this study provides significant contributions by empirically investigating the root causes of the interplay between mobility changes, pandemic severity, and lockdown regimes in 88 countries in the early stage of the pandemic. To perform this study, data were collected from multiple sources. An SEM was developed to find out the complex relationships among the observed variables and latent dimensions. Results from the SEM exhibit that different lockdown measures have significant repercussions to maintain social distancing. However, pandemic severity and socioeconomic and institutional context factors have no significant impact to sustain social distancing practices. The results also explain that socioeconomic and institutional context factors have significant effects on increasing pandemic severity. Elderly people, globalization, and employment in the service sector are primarily responsible for a higher number of coronavirus infection cases and deaths in many countries (e.g., USA, UK, Italy, and Spain). Moreover, social distancing is reasonably able to reduce the severity of coronavirus pandemic, although the impacts are marginal at the granularity of national populations (−0.162). It is also understood that lockdown measures affect the socio-economic context of the countries along with reducing the severity of the coronavirus pandemic. People are adjusting their lifestyle and travel pattern to cope with the new circumstances. Individuals and industries are adopting new alternative strategies to keep pace with the global trend. New possibilities are emerging in the world (i.e., a greater use of information and communication technologies in business and personal life) to flourish in a new environment. Thus, it is surmised that this new normal situation is shaping the personal and business world in such a way to keep moving and meet every demand of the people.

Several policy implications can be drawn from this analysis. An effective way to maintain social distancing is to implement strict lockdown measures. Rather than putting into effect casual stay-at-home recommendations and piecemeal efforts, comprehensive and strict lockdown measures are indispensable to maintain social distancing that can reduce coronavirus infection cases and deaths in a meaningful way. However, since globalization is a reality in the modern era, imposing strict restrictions on people and freight movement within and outside country boundaries is detrimental to the economy and to business partnership. Thus, alternative strategies (e.g., e-shopping, application of information technology) should be undertaken by the authorities to ensure the safe transfer of the people and freight from origin to destination and continue international trade during crisis times.

Despite making significant and timely contributions, the strengths of this study are bound by a few cautionary remarks. First, the Google mobility report was prepared based on data collected from Google account users who turned on their travel location history setting [38]. Thus, it may not represent the true travel behaviors of the general population. More generally, the sway that data quality may have on the conclusions of the analysis should not be brushed off. Data quality may be a concern because health outcome variables are difficult to measure with good accuracy as the pandemic unfolds. Furthermore, international studies are notoriously difficult to conduct due to the heterogeneous adherence to data quality standards in different national contexts. Second, data were collected from multiple sources and integrated to perform the analysis. Thus, it is very challenging to make consensus and consistent policy decisions that can be applied generally. Thirdly, to deal with the ambiguous definition of trips, a comparative analysis of essential versus non-essential travels can be performed based on a recent dataset on changes in the visits to non-essential venues (e.g., restaurants, department stores, and cinemas) published by Unacast [35]. Finally, this study has been conducted at the coarse geographic resolution of countries. Thus, a future study at a finer scale would provide more insights on the interplay between social distancing, pandemic severity, and lockdown regimes. In addition, we propose to pursue further research at the interface of mobility changes, pandemic severity, and lockdown regimes as the COVID-19 pandemic continues to afflict populations around the globe. As more complete time series become available and as the pandemic will have eased into other phases, the stability of our model, or alternatively its dynamic properties, will be critically important to assess to better prepare the world for future pandemics under changing socio-politico-medical contexts. However, this horizon maybe 6, 9, or even 12 months away. We believe the present analysis and results achieved here have value as they stand, as they capture the reality of the pandemic a few months after its global onset.

**Author Contributions:** Data curation, M.M.R., J.-C.T. and K.C.P.; Formal analysis, M.M.R., J.-C.T. and K.C.P.; Investigation, M.M.R., J.-C.T. and K.C.P.; Methodology, M.M.R. and J.-C.T.; Software, M.M.R.; Supervision, J.-C.T.; Validation, J.-C.T.; Visualization, M.M.R. and K.C.P.; Writing—original draft, M.M.R., J.-C.T. and K.C.P.; Writing—review & editing, M.M.R., J.-C.T. and K.C.P.   All authors have read and agreed to the published version of the manuscript.

**Funding:** This research did not receive any specific funding from agencies in public, private, and non-profit organizations.

**Conflicts of Interest:** The authors declare no conflict of interest. The authors are responsible for the contents of this paper.

**Appendix A. List of Variables Tested in the Model**

Table 1. Description of the variables considered in the study.

| Variable | Description | Measure | Source |
|---|---|---|---|
| l_case | Total infection cases per 1 million population | Cases/1M | Worldometer |
| l_death | Total deaths per 1 million population | Deaths/1M | Worldometer |
| Case_Mar20 | Total number of infection cases on March 20 | # | Worldometer |
| Case_May15 | Total number of infection cases on May 15 | # | Worldometer |
| Wcase_Pre | Weekly change of infection cases before April 17 | # | Worldometer |
| Wcase_post | Weekly change of infection cases after April 17 | # | Worldometer |
| Death_Mar20 | Total number of deaths on March 20 | # | Worldometer |
| Death_May15 | Total number of deaths on May 15 | # | Worldometer |
| Wdeath_Pre | Weekly change of death counts before April 17 | # | Worldometer |
| Wdeath_post | Weekly change of death counts after April 17 | # | Worldometer |
| RR | Percentage change of mobility to retail and recreation POIs compared to baseline | % | Google |
| TS | Percentage change of mobility to transit stations POIs compared to baseline | % | Google |
| WP | Percentage change of mobility to workplaces compared to baseline | % | Google |
| RD | Percentage change of time spent at home compared to baseline | % | Google |
| PODEN | Population density | Pop/km2 | UN |
| MFR | Male-female ratio | Ratio | UN |
| GDPGR | Annual GDP growth rate | % | UN |
| GDPP | GDP per capita | $ | UN |
| AE | Percentage of population employed in agriculture | % | UN |
| IE | Percentage of population employed in industry | % | UN |
| SE | Percentage of population employed in services | % | UN |
| UR | Percentage of unemployed people in the workforce | % | UN |
| CPI | Consumer Price Index | Index | UN |
| HE | Percentage of health expenditure to total GDP | % | UN |
| HP | Number of physicians per 1,000 population | # | UN |
| FA | Percentage of forested areas to country land area | % | UN |
| Tourist | Number of tourist/visitor arrivals at national borders (000) | # | UN |
| EI | Average years of schooling (adults) and expected years of schooling (children) | Index | UNDP |
| LR | Percentage of population aged 15 and above who can read and/or write | % | UNDP |
| ExIm | Percentage of exports and imports of goods and services to total GDP | % | UNDP |
| FDI | Percentage of additional long-term and short-term capital of total GDP | % | UNDP |
| Remit | Percentage of earning by migrants to total GDP | % | UNDP |
| IU | Percentage of people with access to the internet | % | UNDP |
| MU | Mobile phone subscriptions per 100 people | # | UNDP |
| MR | Ratio of difference between in-migrants and out-migrants to the average population per 1,000 people | Ratio | UNDP |
| MA | Median age of the people | Year | UNDP |
| DR | Number of elderly (65 and older) per 100 working age (15–64) | Ratio | UNDP |
| UR | Percentage of people living in urban areas | % | UNDP |
| Temp | Average temperature in April (DC: Degree Centigrade) | DC | Weatherbase |
| Rainfall | Average precipitation in April | Mm | Weatherbase |
| DE1 | Days elapsed since the lockdown to April 17 | Days | Different sources |
| DE2 | Days elapsed between first case and imposing social distancing measures | Days | Different sources |
| PDI | Power Distance Index | Index | Hofstede |
| IDV | Individualism Versus Collectivism | Index | Hofstede |
| MAS | Masculinity Versus Feminity | Index | Hofstede |
| UAI | Uncertainty Avoidance Index | Index | Hofstede |
| LTO | Long Term Orientation versus Short Normative Orientation | Index | Hofstede |
| IVR | Indulgence Versus Restraint | Index | Hofstede |
| EFS | Economic Freedom Score | Score | The Fraser Institute |
| KOFGI | KOF Globalisation Index (2017) | Index | KOF Swiss Economic Institute |
| Regime | Regime classification (2008) | Type | Xavier Marquez |
| Agereg | Age in years of the current regime as classified by regime (2008) | Year | Xavier Marquez |
| FC | Date of the first reported infection case | Date | Worldometer |
| LRL | Localized recommended lockdown (1 = Yes, 0 = No) | Dummy (1, 0) | BBC |
| LL | Localized lockdown (1 = Yes, 0 = No) | Dummy (1, 0) | BBC |
| NRL | National recommended lockdown (1 = Yes, 0 = No) | Dummy (1, 0) | BBC |
| NL | National lockdown (1 = Yes, 0 = No) | Dummy (1, 0) | BBC |
| Asia | Country located in Asia continent (1 = Yes, 0 = No) | Dummy (1, 0) | Wikipedia |
| Africa | Country located in Africa continent (1 = Yes, 0 = No) | Dummy (1, 0) | Wikipedia |
| Europe | Country located in Europe continent (1 = Yes, 0 = No) | Dummy (1, 0) | Wikipedia |
| North America | Country located in North America continent (1 = Yes, 0 = No) | Dummy (1, 0) | Wikipedia |
| South America | Country located in South America continent (1 = Yes, 0 = No) | Dummy (1, 0) | Wikipedia |
| Australia | Country located in Australia continent (1 = Yes, 0 = No) | Dummy (1, 0) | Wikipedia |
| ME | Country located in Middle East (1 = Yes, 0 = No) | Dummy (1, 0) | Wikipedia |
| SC | Closing of schools and universities (1 = required closing, 0 = no measure/recommended closing) | Dummy (1, 0) | Oxford University |
| WPC | Closing of workplaces (1 = required closing, 0 = no measure/recommended closing) | Dummy (1, 0) | Oxford University |
| CPE | Cancellation of public events (1 = required canceling, 0 = no measure/recommended canceling) | Dummy (1, 0) | Oxford University |
| RG | Restriction on gatherings (1 = restriction on gathering-less than 100 people, 0 = no restrictions/restriction on very large gathering-above 100 people) | Dummy (1, 0) | Oxford University |
| CPT | Closing of public transport (1 = required closing, 0 = no measure/recommended closing) | Dummy (1, 0) | Oxford University |
| SH | Stay-at-home order (1 = required not to leave, 0 = no measure/recommended not to leave) | Dummy (1, 0) | Oxford University |
| RIM | Restriction on internal movement between cities/regions (1 = required not to travel, 0 = no measure/recommended not to travel) | Dummy (1, 0) | Oxford University |
| RIT | Restriction on international travel (1 = ban/quarantine arrivals, 0 = no restriction/screening arrivals) | Dummy (1, 0) | Oxford University |
| CP | Direct cash payments to people who lose their jobs or cannot work (1 = income support from govt., 0 = no income support | Dummy (1, 0) | Oxford University |
| FO | Freezing financial obligations for households (1 = debt/contract relief, 0 = no measures) | Dummy (1, 0) | Oxford University |
| ES | Announced economic stimulus spending | $ | Oxford University |
| PIC | Public campaigns (1 = public information campaign, 0 = public officials urging caution) | Dummy (1, 0) | Oxford University |
| TP | Testing policy (1 = comprehensive testing policy, 0 = no/limited testing policy) | Dummy (1, 0) | Oxford University |
| CT | Contact testing after positive diagnosis (1 = comprehensive testing policy, 0 = no/limited testing policy) | Dummy (1, 0) | Oxford University |
| EIH | Short term spending on healthcare system (e.g., hospitals, masks) | $ | Oxford University |
| IV | Public spending on Covid-19 vaccine development | $ | Oxford University |
| SI | Government Response Stringency Index (0 to 100, 100 = strictest) | Index | Oxford University |
| AGE65 | Percentage of population aged 65 and older | % | Oxford University |
| AGE70 | Percentage of population aged 70 and older | % | Oxford University |
| DB | Percentage of people with diabetes | % | Oxford University |
| FS | Percentage of female smokers | % | Oxford University |
| MS | Percentage of male smokers | % | Oxford University |
| HB | Hospital beds per 1000 people | Beds/1k people | Oxford University |
| PD | Presidential democracy | Dummy (1, 0) | Oxford University |
| SPD | Mixed (semi-presidential) democracy | Dummy (1, 0) | Oxford University |
| ParD | Parliamentary democracy | Dummy (1, 0) | Oxford University |
| CD | Civilian dictatorship | Dummy (1, 0) | Oxford University |
| MD | Military dictatorship | Dummy (1, 0) | Oxford University |
| RD | Royal dictatorship | Dummy (1, 0) | Oxford University |

**Publisher's Note:** MDPI stays neutral with regard to jurisdictional claims in published maps and institutional affiliations.